\documentclass[
aps,prl,showpacs,superscriptaddress,numerical,amsmath,amssymb,floatfix,reprint]{revtex4-1}
\usepackage[T1]{fontenc}
\usepackage[utf8]{inputenc}
\usepackage[
pdffitwindow=true,
colorlinks=true,
frenchlinks=false,
linkcolor=blue,
anchorcolor=blue,
citecolor=blue,
filecolor=blue,
urlcolor=blue,
bookmarks=true,
bookmarksopen=true,
bookmarksnumbered=true,
bookmarksopenlevel=1,
plainpages=false,
pdfpagelayout=TwoPageLeft,
pdfpagelabels=true,
breaklinks]{hyperref}
\usepackage[main=english]{babel}
\usepackage[dvipsnames]{xcolor}
\usepackage[per-mode=symbol,separate-uncertainty]{siunitx}
\usepackage{graphicx}
\usepackage{dcolumn}
\usepackage{color}
\usepackage{graphicx,wrapfig,lipsum}
\usepackage{bm}

\usepackage{color}
\usepackage{ulem}
\usepackage{float}
\usepackage{chemformula}

\begin{document}
	\title{Quantum microwave parametric interferometer
	}
	% Force line breaks with \\
	
\author{F.\,Kronowetter}
\email[]{fabian.kronowetter@wmi.badw.de}
\affiliation{Walther-Mei{\ss}ner-Institut, Bayerische Akademie der Wissenschaften, 85748 Garching, Germany} 
\affiliation{School of Natural Sciences, Technische Universität München, 85748 Garching, Germany}
\affiliation{Rohde \& Schwarz GmbH \& Co. KG, 81671 Munich, Germany}

\author{F.\,Fesquet}
\affiliation{Walther-Mei{\ss}ner-Institut, Bayerische Akademie der Wissenschaften, 85748 Garching, Germany}
\affiliation{School of Natural Sciences, Technische Universität München, 85748 Garching, Germany}

\author{M.\,Renger}
\affiliation{Walther-Mei{\ss}ner-Institut, Bayerische Akademie der Wissenschaften, 85748 Garching, Germany}
\affiliation{School of Natural Sciences, Technische Universität München, 85748 Garching, Germany}

%\author{M.\,Handschuh}
%\affiliation{Walther-Mei{\ss}ner-Institut, Bayerische Akademie der Wissenschaften, 85748 Garching, Germany}
%\affiliation{Technical University of Munich, TUM School of Natural Sciences, Physics Department, 85748 Garching, Germany}

\author{K.\,Honasoge}
\affiliation{Walther-Mei{\ss}ner-Institut, Bayerische Akademie der Wissenschaften, 85748 Garching, Germany}
\affiliation{School of Natural Sciences, Technische Universität München, 85748 Garching, Germany}

\author{Y.\,Nojiri}
\affiliation{Walther-Mei{\ss}ner-Institut, Bayerische Akademie der Wissenschaften, 85748 Garching, Germany}
\affiliation{School of Natural Sciences, Technische Universität München, 85748 Garching, Germany}

\author{K.\,Inomata}
\affiliation{RIKEN Center for Quantum Computing (RQC), Wako, Saitama 351-0198, Japan}
\affiliation{National Institute of Advanced Industrial Science and Technology, 1-1-1 Umezono, Tsukuba, Ibaraki, 305-8563, Japan}

\author{Y.\,Nakamura}
\affiliation{RIKEN Center for Quantum Computing (RQC), Wako, Saitama 351-0198, Japan}
\affiliation{Department of Applied Physics, Graduate School of Engineering, The University of Tokyo, Bunkyo-ku, Tokyo 113-8656, Japan}

\author{A.\,Marx}
\affiliation{Walther-Mei{\ss}ner-Institut, Bayerische Akademie der Wissenschaften, 85748 Garching, Germany}

\author{R.\,Gross}
\affiliation{Walther-Mei{\ss}ner-Institut, Bayerische Akademie der Wissenschaften, 85748 Garching, Germany}
\affiliation{School of Natural Sciences, Technische Universität München, 85748 Garching, Germany}
\affiliation{Munich Center for Quantum Science and Technology (MCQST), 80799 Munich, Germany}

\author{K.\,G.\,Fedorov}
\email[]{kirill.fedorov@wmi.badw.de}
\affiliation{Walther-Mei{\ss}ner-Institut, Bayerische Akademie der Wissenschaften, 85748 Garching, Germany}
\affiliation{School of Natural Sciences, Technische Universität München, 85748 Garching, Germany}

	\date{\today}% It is always \today, today,
	 %Before starting of \begin{abstract}	
	\begin{abstract}
 %Classical interferometers are indispensable tools for the precise determination of various physical quantities. Their accuracy is bound by the standard quantum limit. This limit can be overcome by using quantum states or nonlinear quantum elements. Such nonlinear quantum interferometers have been thoroughly investigated at optical frequencies, while leaving the microwave domain largely unexplored. Meanwhile, quantum microwave sensing has developed into a novel and quickly growing research field, which offers exciting perspectives for applications and fundamental studies. Here, we present the experimental study of a nonlinear Josephson interferometer operating in the microwave regime. Our quantum microwave parametric interferometer (QUMPI) consists of two cryogenic $\SI{180}{\degree}$ hybrid ring beam splitters combined with superconducting flux-driven Josephson parametric amplifiers. We perform a systematic analysis of the implemented QUMPI. We find that the Gaussian interferometric power exceeds the shot-noise limit and observe sub-Poissonian photon statistics in its output modes. Furthermore, we identify a low-gain operation regime of the interferometer which is essential for joint measurements in quantum illumination protocols.
 Classical interferometers are indispensable tools for the precise determination of various physical quantities. Their accuracy is bound by the standard quantum limit. This limit can be overcome by using quantum states or nonlinear quantum elements. Here, we present the experimental study of a nonlinear Josephson interferometer operating in the microwave regime. Our quantum microwave parametric interferometer (QUMPI) is based on superconducting flux-driven Josephson parametric amplifiers combined with linear microwave elements. We perform a systematic analysis of the implemented QUMPI. We find that its Gaussian interferometric power exceeds the shot-noise limit and observe sub-Poissonian photon statistics in the output modes. Furthermore, we identify a low-gain operation regime of the QUMPI which is essential for optimal quantum measurements in quantum illumination protocols.
	\end{abstract}
		\maketitle
\textit{Introduction ---} As part of the second quantum revolution, various quantum technologies have matured to a level allowing their use in a plethora of practical applications~\cite{Sibson.2017, Degen.2017}. In particular, the fields of quantum communication, metrology and sensing have made great progress~\cite{Degen.2017, Giovannetti.2006, Hwang.2003,Pogorzalek.2019}. In metrology, the field of interferometry has been widely explored in terms of fundamental physics and resulted in a variety of technical breakthroughs~\cite{Xiao.1987, Schnabel.2017, AbadieJ.etal..2011}. Classical interferometers, such as the Mach-Zehnder interferometer, typically rely on injection of a coherent state into one port of a beam splitter, while only vacuum fluctuations enter the second port~\cite{Born.2013}. Their phase sensitivity is limited by the shot noise of the coherent signal, also known as standard quantum limit~(SQL). The SQL reflects in a $1/\sqrt{N}$ scaling of the phase sensitivity, or equivalently, in a $\sqrt{N}$ scaling of the signal-to-noise ratio (SNR), where $N$ is the photon number of the input coherent state~\cite{Ou.2020, Caves.1981}.
%Although the shot noise scales with $\sqrt{N}$, the signal-to-noise ratio (SNR) scales as $1/\sqrt{N}$ as the measured signal scales linearly with the photon number $N$ of the coherent state~\cite{Caves.1981}.
This linear interferometer sensitivity can be improved by coupling quantum states, such as squeezed states, into the second beam splitter port~\cite{Grangier.1987,Ou.2020}. Alternatively, the SQL can be overcome by making use of nonlinear elements such as parametric amplifiers, leading to interactions between photons~\cite{Giovannetti.2006, Giovannetti.2011, Napolitano.2011, Yurke.1986}. In principle, exploiting the quantum correlations between photons in these states enables achieving the Heisenberg limit~(HL) with a linear scaling of the SNR with respect to~$N$~\cite{Ou.1997, Ou.2020}. While nonlinear interferometers have been investigated at optical frequencies, the microwave domain so far remained largely unexplored due to relatively small energies of microwave photons with frequencies in the 1--10~$\si{\giga\hertz}$ regime and the associated difficulty of single photon detection~\cite{Ou.2020, Yurke.1986, Ou.2012, Hudelist.2014}. Meanwhile, quantum microwave sensing and communication represent novel and rapidly growing fields, which promise groundbreaking fundamental experiments and applications~\cite{Fedorov.2021,Barzanjeh.2015,Flurin.2012,Assouly.2022}.

\begin{figure}[t]
        	\centering
        	\includegraphics{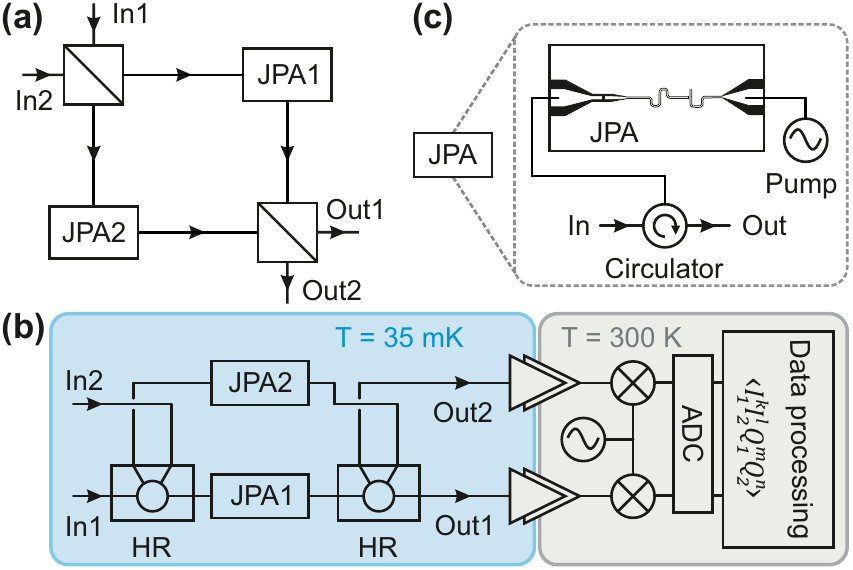}
        	\caption{(a) General scheme of the QUMPI. (b) Details of the experimental setup consisting of a $180\si{\degree}$\,hybrid ring~(HR), which splits and symmetrically superimposes two incoming signals from ports In1 and In2, two JPAs for phase-sensitive amplification, and a second $180\si{\degree}$\,HR, which completes the nonlinear interferometer. Output two-mode signals are detected with a heterodyne microwave receiver and digitally processed to extract statistical signal moments. The latter enable a full state tomography, including the reconstruction of the displacement vector and covariance matrix. (c) A circulator separates the incoming and outgoing signals for each JPA.}
        	\label{fig:Fig_1}
            \end{figure}

In this Letter, we present an experimental realization of a nonlinear microwave interferometer making use of Josephson-junction-based superconducting quantum circuits (see Fig.~\ref{fig:Fig_1}). This quantum microwave parametric interferometer~(QUMPI) consists of two linear balanced microwave beam splitters and two active quantum devices in the form of flux-driven Josephson parametric amplifiers~(JPAs). We experimentally characterize the QUMPI by injecting various Gaussian states and analyze its performance by comparing experimental results to predictions of an input--output theory model in terms of output photon numbers, interferometric power, and second-order correlation functions. We observe that the interferometric power of the QUMPI exceeds the SQL, which highlights the potential of our scheme in precision metrology~\cite{Giovannetti.2011}. With symmetric coherent signal inputs, our interferometer shows photon anti-bunching between the outputs. This is captured by a second-order cross-correlation function, $g^{(2)}_\mathrm{C} < 1$, and reflects the nonclassical nature of the QUMPI. For the specific operating point with equal phase-sensitive gain amplitudes, $G_1=G_2=G$, of the JPAs and orthogonal amplification angles, $\gamma_1 = \gamma_2 + \pi/2$, the input--output operator relation of the circuit coincides with that of a SU(1,1)~interferometer~\cite{Ou.2020}. The QUMPI can be also used for analog Bell measurements in microwave quantum teleportation \cite{Fedorov.2021}. In the low-gain operating regime, $G \gtrsim 1$, the QUMPI realizes an effective two-mode, phase-conjugate signal mixing, which is an integral part of joint receivers for quantum radar schemes~\cite{LasHeras.2017,Guha.2009,Lloyd.2008, Sorelli.2022}. All these findings demonstrate the practical versatility and fundamental potential of the considered scheme.
            
\textit{Experiment ---} In Fig.\,\ref{fig:Fig_1}(a), we present the idea of the QUMPI. Input signals at ports In1 and In2 are split and subsequently fed into JPA1 and JPA2. Then, the nonlinearly amplified signals from the JPAs interfere and leave the circuit at ports Out1 and Out2. Figure~\ref{fig:Fig_1}(b) shows a detailed circuit layout of our experiment. We employ two symmetric hybrid rings (HRs) as microwave beam splitters and two superconducting flux-driven JPAs. The latter are operated at a frequency of $\omega _0 / 2\pi = \SI{5.48}{\giga\hertz}$. Experimental bandwidths of JPA1 and JPA2 are 2\,$\si{\MHz}$ and 10\,$\si{\MHz}$, respectively. Low-loss microwave cryogenic circulators are used to separate the incoming and outgoing signals of the JPAs~[cf. Fig.\,\ref{fig:Fig_1}(c)]. We operate both JPAs in the phase-sensitive amplification regime by pumping them at twice the resonance frequency, $\omega_\mathrm{p}=2\omega_0$~\cite{Yurke.1989,Yamamoto.2008}. The microwave interferometer arms are tailored to have identical lengths with an accuracy of \SI{1}{\milli\metre}. At the carrier frequency of around \SI{5.5}{\giga\hertz}, the corresponding wavelength in our superconducting cable is approximately~\SI{38}{\milli\metre}. Given the interferometer arm accuracy in comparison with the signal wavelength, microwave signals traveling along different interferometer paths do not acquire a significant relative, path-induced phase shift. However, an overall phase difference also depends on JPA-induced phase shifts. These can be adjusted by fine-tuning the JPA operation frequency with an external magnetic flux~\cite{Yamamoto.2008}. The output state tomography relies on heterodyne measurements with an FPGA-based digitization setup~\cite{Menzel.2012,Eichler.2011}. After digital down-conversion and filtering, we use a reference-state reconstruction method to extract statistical field quadrature moments and reconstruct a covariance matrix of quantum states at a certain reference point~\cite{Menzel.2012,Eichler.2011,Renger.2022}. This reference point can be defined by performing a Planck spectroscopy which provides a precise \textit{in-situ} photon number calibration of the output lines~\cite{Mariantoni.2010}.
			\begin{figure}[t]
    	    \centering
	        \includegraphics{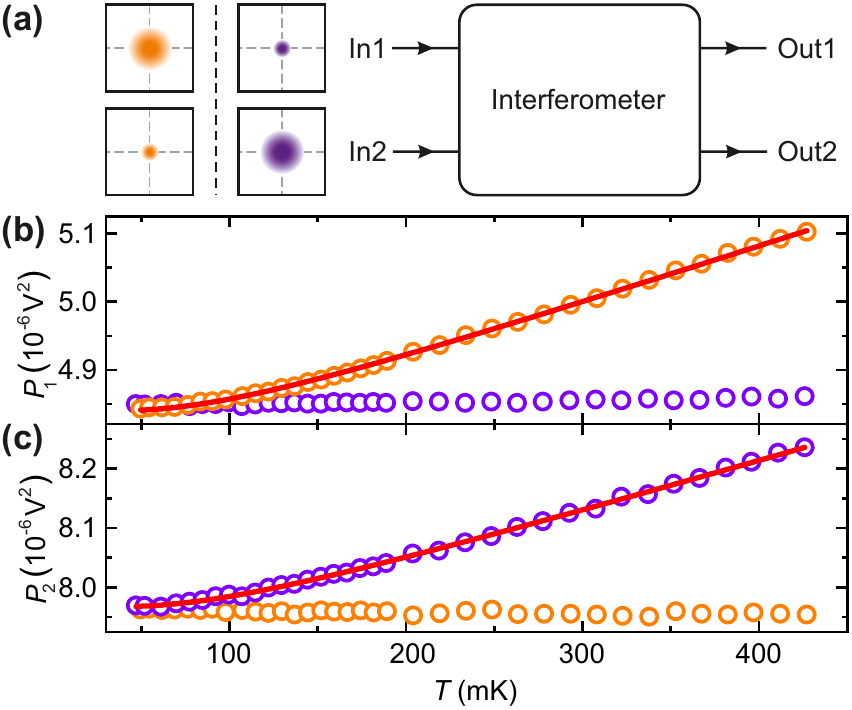}
    	    \caption{Planck spectroscopy of the interferometer in the linear regime for output channel powers (b)~$P_1$ at Out1 and (c)~$P_2$ at Out2. Orange points correspond to thermal-state injection at In1 and vacuum at In2, purple points correspond to the inverted case of thermal state injection at In2 and vacuum state at In1, as depicted schematically in the quadrature planes in~(a). The temperature dependence of $P_1$~($P_2$) for orange~(purple) data points yields the photon number calibration for the interferometer and verifies its functionality. The corresponding error bars are smaller than the symbol size. The solid red lines represent fits based on Planck's law.}
	        \label{fig:Fig_2}
            \end{figure}
            
\textit{Results and Discussion ---} A systematic study of the QUMPI requires careful calibration and pre-characterization. In particular, a precise and stable control of the JPAs is the prerequisite for subsequent measurements and analysis of the interferometer. First, we detune both JPAs from the intended operation frequency and switch our interferometer into the linear regime. This detuning is implemented by changing an external magnetic field generated by superconducting coils mounted on top of each JPA. We perform Planck spectroscopy of our system by injecting a thermal state generated by a heatable attenuator at one input and vacuum at the other, as illustrated in Fig.\,\ref{fig:Fig_2}(a)~\cite{Mariantoni.2010}. Figures~\ref{fig:Fig_2}(b) and~\ref{fig:Fig_2}(c) show the experimental results of these measurements. We observe both constructive and destructive interference of the broadband thermal signals, as expected in a symmetric linear interferometer. Thermal-signal injection at In1 (orange points) results in the temperature dependence of $P_1$, while $P_2$ remains independent of $T$ due to destructive interference. The inverted case of thermal signal injection at In2 (purple points) demonstrates the system symmetry, which is reflected in $P_2(T)$ and a $T$-independent response of $P_1$.
\begin{figure}[t]
	\centering
	        \includegraphics{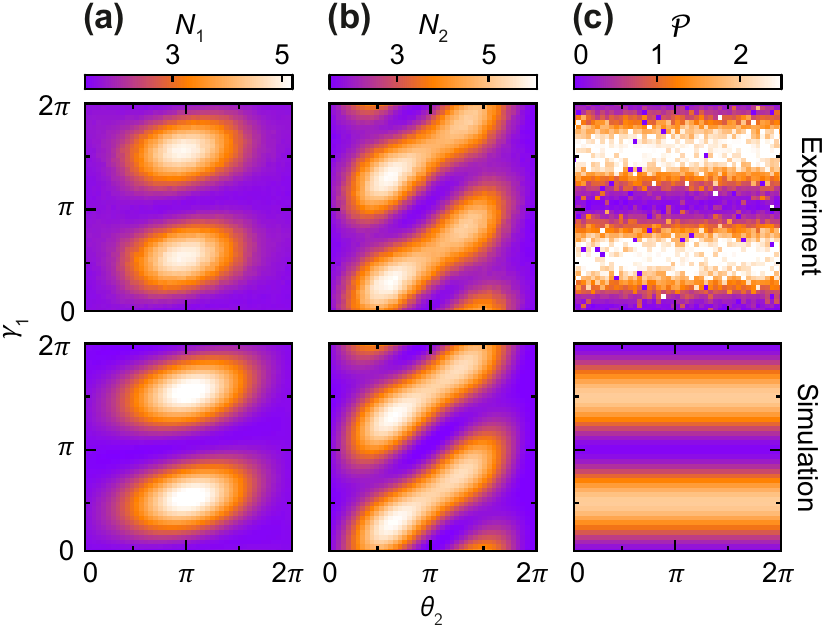}
	        \caption{Interferometer measurements with coherent signals applied to In1 and In2. The corresponding displacement amplitudes are $|\alpha_1|=|\alpha_2|=0.83(5)$, and the displacement angle $\theta_1$ is fixed to $0.64\pi$, while $\theta_2$ varies from 0 to 2$\pi$. Both JPAs are operated as squeezers with the average gain $\overline{G}_{1,2}=\SI{7.73}{dB}$ and squeezing angle $\gamma_1$ varying from 0 to 2$\pi$, while $\gamma_2=0$. Top row shows the experimentally reconstructed photon numbers (a)~$N_1$ and (b)~$N_2$ at the ports Out1 and Out2, respectively, as a function of $\theta_2$ and $\gamma_1$. (c) Gaussian interferometric power $\mathcal{P}$ of the studied circuit illustrating the two-mode state probe capabilities.}
	        \label{fig:Fig_3}
\end{figure}

As a second part of the calibration routine, we tune both JPAs to the same resonance frequency of $\omega _0 / 2\pi = \SI{5.48}{\giga\hertz}$. This step converts the QUMPI into the nonlinear regime. The JPA response is controlled via magnitude and phase of the coherent pump tone. Since we expect the best interferometric performance for a balanced gain of $G_1=G_2$~\cite{Ou.2020}, we inject vacuum states at both circuit inputs and pump the JPAs with varying magnitude and phase~\cite{ Menzel.2012}. We balance the produced two-mode states at the interferometer output by minimizing asymmetries of the local output modes. These asymmetries result in squashed variances $\sigma_{\mathrm{s,}i}^2$ and amplified variances $\sigma_{\mathrm{a,}i}^2$ ($i=1,2$ denotes path 1 and path 2, respectively) of the local phase space distributions. We define a balancing criterion as $\mathcal{B}=\left(\sigma_{\mathrm{s,}1}^2/\sigma_{\mathrm{a,}1}^2\right)\cdot\left(\sigma_{\mathrm{s,}2}^2/\sigma_{\mathrm{a,}2}^2\right)$. Note that $\mathcal{B}$ is unity for an ideal balanced state and decreases with increasing imbalance between the local variances. For our system, we observe $\mathcal{B}$ reaching values of around 0.91, close to the optimum ($\mathcal{B} =1$)~\cite{Kronowetter.2023}. Finite asymmetries and insertion losses of the HRs, as well as the nonidentical noise properties of the JPAs, limit the balancing.

After the calibration, we can investigate the nonlinear interferometer response to coherent signals applied to both input ports, In1 and In2, with a photon number of $|\alpha_1|^2 = |\alpha_2|^2=0.69(7)$, where $|\alpha_i|$~($i =1,2$) are the respective displacement amplitudes~\cite{Fedorov.2016}. We fix one coherent displacement angle, $\theta_1=0.64\pi$, while varying the other, $\theta_2$, from 0~to~2$\pi$. Both JPAs are operated in the phase-sensitive amplification regime with an average gain $\overline{G}_{1,2}=\SI{7.73}{dB}$ and an average number $\overline{n}_{1,2}=0.238$ of added noise photons referred to the JPA inputs. The JPA2 squeezing angle, $\gamma_2$, is fixed to 0 and we vary $\gamma_1$ from 0~to~2$\pi$. We compare the acquired data with a theoretical model of our system based on the input--output formalism. This model encompasses losses of the different components of the experimental setup (see Ref.\,\cite{Kronowetter.2023} for more details). The JPA noise, JPA gain and squeezing angle, phase and number of injected coherent photons are calculated from the reconstructed displacement vectors and covariance matrices of the recorded signal quadratures~\cite{Menzel.2012,Eichler.2011}. The only free parameter in our model is a different phase acquired along the two interferometer arms. Figures~\ref{fig:Fig_3}(a) and~\ref{fig:Fig_3}(b) show the photon numbers $N_i=\langle \hat{a}_i^{\dagger}\hat{a}_i\rangle$ at the respective outputs Out$i$. Here, $\hat{a}_i^\dagger$~($\hat{a}_i$) is the photon creation~(annihilation) operator. The bottom row of Fig.\,\ref{fig:Fig_3} shows the model prediction. The common color bars for each column underline a good agreement between experiment and theory. Since our model intrinsically corresponds to a nonlinear interferometer, this agreement confirms that our system acts as such. The asymmetry in the patterns between Fig.~\ref{fig:Fig_3}(a) and ~\ref{fig:Fig_3}(b) stems from the nonlinear character of our interferometer.

Next, we evaluate the interferometric power~(IP) of the QUMPI. For a bipartite quantum probe state, the IP defines the worst-case precision of a parameter estimation, where the corresponding parameter experiences unitary dynamics in one of the two subsystems (e.g., a phase shift of the signal in one arm of the interferometer)~\cite{Adesso.2014}. The respective IP, $\mathcal{P}$, is defined as
\begin{equation}
    \mathcal{P} \left(\rho_\mathrm{AB} \right) = \frac{1}{4}\,\inf_{\hat{U}_\mathrm{A}}\mathcal{F}\left( \rho^{\Phi,\hat{U}_\mathrm{A}}_\mathrm{AB} \right) ,
\end{equation}
where $\rho_\mathrm{AB}$ is the two-mode probe state, $\hat{U}_\mathrm{A}$ is an arbitrary unitary transformation of the subsystem A, $\mathcal{F}$ is the quantum Fisher information, and $\Phi$ is the corresponding estimator~\cite{Adesso.2014,Girolami.2014}. Remarkably, the IP provides a measure of bipartite discord-type correlations for Gaussian states beyond pure entanglement~\cite{Modi.2012,Adesso.2014}. We apply the expressions from Ref.\,\cite{Adesso.2014} to our theory model, as well as to the reconstructed experimental covariance matrices, in order to extract the IP of the QUMPI. Figure~\ref{fig:Fig_3}(c) depicts both the experimental and theoretical IP as a function of $\theta_2$ and $\gamma_1$. The data in Fig.~\ref{fig:Fig_3}(c) is independent of $\theta_2$, since $\mathcal{P}$ is invariant under local unitary operations~\cite{Adesso.2014}. Furthermore, $\mathcal{P}$ goes to zero for parallel amplification angles $\gamma_1=\gamma_2+n\pi$ ($n=1,2,...$), where output states become separable, and is maximal for orthogonal amplification, where the states are entangled. In this context, the SQL sets an upper bound $\mathcal{P}_\mathrm{SQL} = N$ for separable two-mode probe states, where $N$ is the mean photon number in the probing subsystem. Pure two-mode squeezed states saturate the HL with $\mathcal{P}_\mathrm{HL} = N(N+1)$~\cite{Adesso.2014}. The simulated IP qualitatively reproduces the measurement data, but the maximum theoretical amplitude is smaller, as it can be seen from Fig.~\ref{fig:Fig_3}(c). We attribute this deviation to possible misestimates of losses in the underlying photon number calibration. Both, maximum theoretical ($\mathcal{P}_\mathrm{theory}/\mathcal{P}_\mathrm{SQL}=1.38$) and experimental ($\mathcal{P}_\mathrm{exp}/\mathcal{P}_\mathrm{SQL}=1.70$) values exceed the SQL but do not reach the HL ($\mathcal{P}_\mathrm{theory}/\mathcal{P}_\mathrm{HL}=0.58$, $\mathcal{P}_\mathrm{exp}/\mathcal{P}_\mathrm{HL}=0.65$). The presence of finite noise in the system prevents reaching an IP closer to the HL. We note, however, that $\mathcal{P}_\mathrm{exp} > \mathcal{P}_\mathrm{SQL}$ provides direct evidence that the QUMPI exceeds the $\sqrt{N}$ scaling of the SNR, since $\mathrm{SNR}\propto\sqrt{\mathcal{P}}$.

            \begin{figure}[t]
	        \centering
    	    \includegraphics{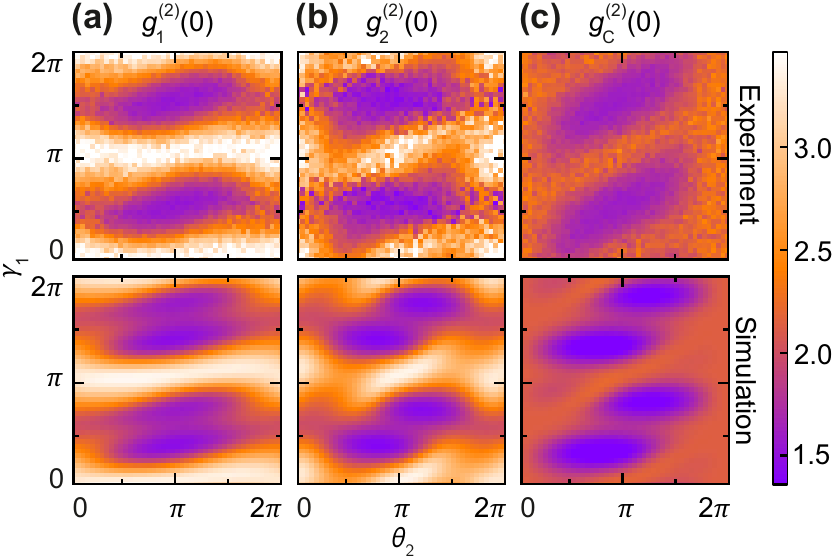}
	        \caption{Second-order correlation analysis of the QUMPI. Single-mode second-order correlation function, $g_1^{(2)} (0)$ and $g_2^{(2)} (0)$, at the interferometer ports (a)~Out1 and (b)~Out2. (c)~Second-order cross-correlation function, $g_\mathrm{C}^{(2)} (0)$, between ports Out1 and Out2. The experimental parameters are identical to those in Fig.\,\ref{fig:Fig_3}.}
	        \label{fig:Fig_4}
            \end{figure}

In order to study correlation properties and related intensity fluctuation statistics of the QUMPI, we analyze the zero-delay-time second-order correlation function, $g^{(2)} (0)$, for the single-mode fields at the interferometer outputs, as well as the cross-correlations between the outputs~\cite{Fedorov.2018}. For the two-mode fields, the respective auto-correlation function, $g_i^{(2)} (0)$, can be written as
\begin{equation}
\label{eq:g2auto}
g_i^{(2)} (0) = \frac{\langle \hat{a}_i^\dagger \hat{a}_i^\dagger \hat{a}_i \hat{a}_i\rangle}{\langle \hat{a}_i^\dagger \hat{a}_i\rangle^2}, 
\end{equation}
where $i=1,2$. The associated second-order cross-correlation function, $g_\mathrm{C}^{(2)} (0)$, can be expressed as
\begin{equation}
\label{eq:g2cross}
g_\mathrm{C}^{(2)} (0) = \frac{\langle \hat{a}_1^\dagger \hat{a}_1^\dagger \hat{a}_1 \hat{a}_1\rangle + \langle \hat{a}_2^\dagger \hat{a}_2^\dagger \hat{a}_2 \hat{a}_2\rangle + 2\langle \hat{a}_1^\dagger \hat{a}_1 \hat{a}_2^\dagger \hat{a}_2\rangle}{\left( \langle \hat{a}_1^\dagger \hat{a}_1\rangle + \langle \hat{a}_2^\dagger \hat{a}_2\rangle \right) ^2}.
\end{equation}
The experimentally obtained data for $g_1^{(2)} (0)$, $g_2^{(2)} (0)$ and $g_\mathrm{C}^{(2)} (0)$ as a function of $\theta_2$ and $\gamma_1$ are depicted in Fig.\,\ref{fig:Fig_4}. The bottom row shows the respective theoretical predictions~\cite{Olivares.2018, Kronowetter.2023}. Our model reproduces the experimental observations. In accordance with the model, the local output modes show correlation functions indicating photon bunching,  $g_1^{(2)} (0),\,g_2^{(2)} (0),\,g_\mathrm{C}^{(2)} (0)>1$.
            \begin{figure}[t]
    	    \centering
	        \includegraphics{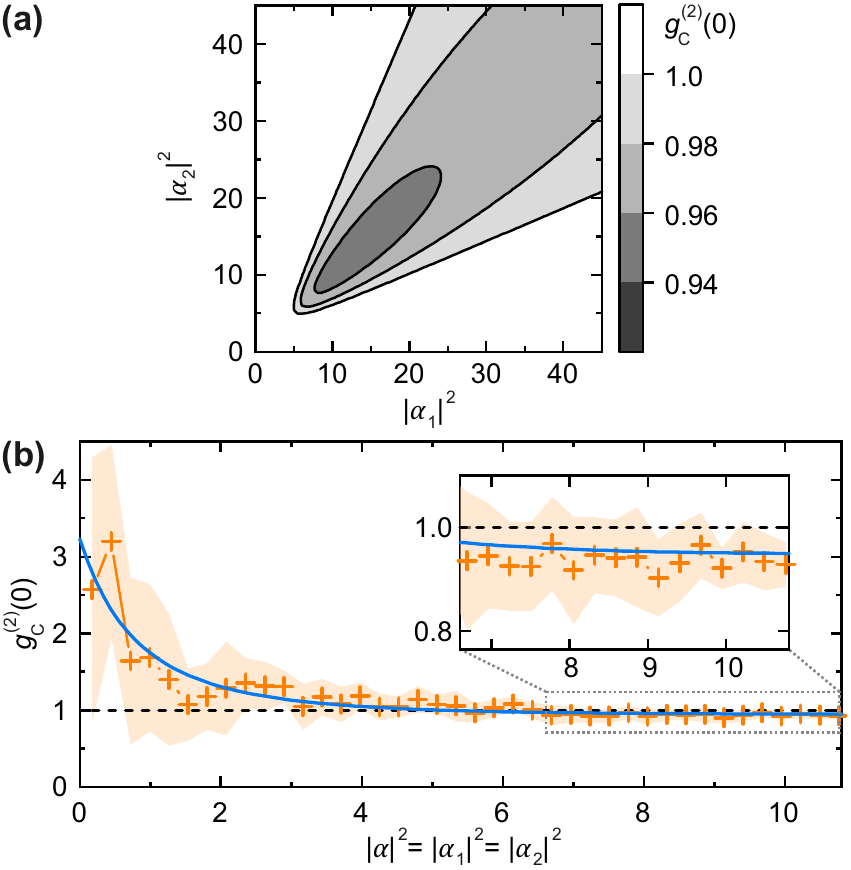}
	        \caption{Intensity cross-correlations, $g_\mathrm{C}^{(2)} (0)$, of the interferometer output fields for variable displacement amplitudes of coherent input signals. (a)~Theoretical model predictions as a function of the number of coherent photons $|\alpha_1|^2$ and $|\alpha_2|^2$ ($\theta_1=\theta_2=0.81\pi$) entering the circuit at In1 and In2, respectively. (b)~Experimental results for $g_\mathrm{C}^{(2)} (0)$ (orange crosses with standard deviation shown in shaded orange) as a function of the symmetrically varied displacement amplitudes. Here, the blue line depicts the theoretical prediction. The black dashed line illustrates the classical limit of $g_\mathrm{C}^{(2)} (0)=1$.}
        	\label{fig:Fig_5}
            \end{figure}

To further explore the QUMPI, we experimentally investigate cross-correlations as a function of the displacement amplitude of the incident coherent states. We observe that for sufficiently large displacement amplitudes, $|\alpha_1|$ and $|\alpha_2|$, and equal displacement angles $\theta_1=\theta_2$ the cross-correlation function $g_\mathrm{C}^{(2)} (0)$ indicates anti-bunching between the interferometer outputs, providing evidence for nonclassical correlations between them~\cite{Olivares.2018}. In Fig.\,\ref{fig:Fig_5}(a), we show $g_\mathrm{C}^{(2)} (0)$ predictions according to our theoretical model. For the experimentally relevant model parameters, most importantly the adapted average JPA gain $\overline{G}_{1,2}=\SI{4.06}{dB}$, Fig.\,\ref{fig:Fig_5}(a) shows that $|\alpha_1|^2,|\alpha_2|^2>5$ is required to realize nonlocal photon anti-bunching, $g_\mathrm{C}^{(2)} (0)<1$~\cite{Kronowetter.2023}. Figure\,\ref{fig:Fig_5}(b) shows the experimental data of $g_\mathrm{C}^{(2)} (0)$ as a function of $|\alpha|^2=|\alpha_{1}|^2 =|\alpha_{2}|^2$. The black dashed line illustrates the classical threshold of $g_\mathrm{C}^{(2)} (0)=1$. The blue line is a cut along the main diagonal of Fig.\,\ref{fig:Fig_5}(a). The inset shows an expanded view of the region where the data points for $g_\mathrm{C}^{(2)} (0)$ drop below the classical limit. Our theory model suggests that the minimal coherent photon number $|\alpha|^2=|\alpha_{1}|^2 =|\alpha_{2}|^2$ to achieve $g_\mathrm{C}^{(2)} (0)<1$ increases with increasing JPA gain, while $g_\mathrm{C,min}^{(2)} (0)$ converges towards unity for large JPA gain. At the same time, $g_\mathrm{C}^{(2)} (0)$ becomes more robust towards noise with increasing JPA gain.

Remarkably, for the specific operating point of the JPAs with equal gain and orthogonal amplification angles, the input--output relations of the QUMPI can be reduced to
\begin{equation}
\label{eq:inout}
\begin{gathered}
\hat{b}_1=\sqrt{G_\mathrm{eff}}\,\hat{a}_1+\sqrt{G_\mathrm{eff}-1}\,\hat{a}_2^\dagger,\\
\hat{b}_2=\sqrt{G_\mathrm{eff}}\,\hat{a}_2+\sqrt{G_\mathrm{eff}-1}\,\hat{a}_1^\dagger,
\end{gathered} 
\end{equation}
with the effective gain $\sqrt{G_\mathrm{eff}}=\mathrm{cosh}(r)$ and the JPA squeezing factor $r=r_1=r_2$ according to the JPA gain $G=\mathrm{exp}(2r)$. The input and output modes are described by bosonic operators $\hat{a}_1, \hat{a}_2$ and $\hat{b}_1, \hat{b}_2$, respectively. The relations in Eq.\,(\ref{eq:inout}) coincide with those describing the so-called Josephson mixer, which can be utilized for producing EPR states of microwave light~\cite{Flurin.2012, Fedorov.2021}. For low effective gain values, $G_\mathrm{eff} \simeq 1$, our interferometer can be applied in a quantum illumination detection scheme for achieving a 3-dB advantage in the error exponent over the ideal classical counterpart~\cite{LasHeras.2017, Guha.2009, Lloyd.2008}. Quantum illumination has been primarily investigated for optical frequencies~\cite{Lopaeva.2013,Zhang.2013,Zhang.2015}, but the recent complete realization of a quantum microwave radar has sparked renewed interest in quantum microwave sensing~\cite{Assouly.2022}. Interestingly, the input--output relations in Eq.\,(\ref{eq:inout}) also coincide with those of a SU(1,1) interferometer, with the exception that the coefficients, $G_\mathrm{eff}$ and $(G_\mathrm{eff} - 1)$, enter linearly in the SU(1,1) implementation~\cite{Ou.2020}. This difference is related to the fact that the parametric amplifiers are connected in series for the conventional SU(1,1) implementation, whereas the JPAs in the QUMPI are arranged in a parallel configuration.

\textit{Conclusion ---} We have realized and systematically analyzed a quantum microwave parametric interferometer. We have performed a detailed investigation of the input-output relations of our QUMPI device with coherent and thermal input states. Our experimental results can be well explained by using a theoretical model based on the input--output quantum formalism. As part of our study, we have demonstrated non-local photon anti-bunching at the QUMPI outputs, characterized by the second-order cross-correlation function, $g_\mathrm{C}^{(2)} (0) < 1$, for coherent input states. The investigated circuit is expected to be useful in many applications ranging from quantum-enhanced interferometry to mode-mixing, as part of a joint quantum receiver in quantum sensing experiments~\cite{LasHeras.2017, Gong.2022}. Furthermore, our findings open a new avenue towards quantum-enhanced nonlinear interferometers in the fast-evolving field of superconducting circuits operating in the microwave regime. Remarkably, current dark matter axion search experiments focus on the frequency range from $\SI{1}{\giga\hertz}$ to $\SI{25}{\giga\hertz}$ and rely on read-out by quantum-limited amplifiers~\cite{Semertzidis.2022}. To this end, the QUMPI could find applications in related dark matter axion search experiments~\cite{Adair.2022}.\\
	
We acknowledge support by the German Research Foundation via Germany’s Excellence Strategy (EXC-2111-390814868), the German Federal Ministry of Education and Research via the project QUARATE (Grant No.13N15380), JSPS KAKENHI (Grant No. 22H04937) and JST ERATO (Grant No. JPMJER1601). This research is part of the Munich Quantum Valley, which is supported by the Bavarian state government with funds from the Hightech Agenda Bayern Plus.

\bibliography{Main}

\end{document}